\documentclass[%
 reprint,
superscriptaddress,
 amsmath,amssymb,
 aps,
pra,
floatfix,
]{revtex4-1}

\usepackage{graphicx}
\usepackage{dcolumn}
\usepackage{bm}
\usepackage{color}
\usepackage[linktocpage,colorlinks=true,linkcolor=blue,citecolor=blue,urlcolor=blue,breaklinks=true]{hyperref}   
\usepackage{MnSymbol}

\newcommand{\ave}[1]{\left\langle #1\right\rangle}

\def\bark{\overline{k}}
\def\ldef{\mathrel{\mathop:}=}
\def\s{\mathbf{s}}
\def\avek{\langle k\rangle}

\def\d{\mathrm{d}}
\def\citeSM#1{ (see SM Section #1)}

\begin{document}
\bibliographystyle{unsrt}

\title{Correlated Growth of Causal Networks}

\author{Jiazhen Liu}
\affiliation{%
Department of Physics, University of Miami, Coral Gables, Florida 33142, USA}%
\author{Kunal Tamang}
\affiliation{%
Department of Physics, University of Miami, Coral Gables, Florida 33142, USA}%
\author{Dashun Wang}
\affiliation{%
Kellogg School of Management, Northwestern University, Evanston, Illinois 60208, USA
}%
\author{Chaoming Song}
\email{c.song@miami.edu}
\affiliation{%
Department of Physics, University of Miami, Coral Gables, Florida 33142, USA}%

\begin{abstract}
The study of causal structure in complex systems has gained increasing attention, with many recent studies exploring causal networks that capture cause-effect relationships across diverse fields. Despite increasing empirical evidence linking causal structures to network topological correlations, the mechanisms underlying the emergence of these correlations in causal networks remain poorly understood. In this work, we propose a general growth framework for causal networks, incorporating two key types of correlations: causal and dynamic. We analytically demonstrate that degree correlations emerge as a consequence of marginal dependencies on these correlations. Our theoretical predictions align quantitatively with empirical data from four large-scale innovation networks. Our theory not only sheds light on the origins of topological correlations but also provides a general framework for understanding correlated growth across causal systems.
\end{abstract}

\maketitle

Causal networks are central to capturing cause-effect relationships in a wide array of fields \cite{jin2019emergence,karrer2009random,thulasiraman2011graphs}. The growing interest in causal networks \cite{pearl1995bayesian,ito2013information}, driven by recent advances in understanding causal structures in various disciplines, including quantum communication \cite{carvacho2017experimental}, cosmology \cite{boguna2014cosmological},  quantum gravity\cite{donoghue2019arrow,hamada2019weak,reall2021causality}, social science \cite{liu2018hot, wang2013quantifying,golosovsky2012stochastic}, computer science \cite{scholtes2014causality}, technologies \cite{jin2019emergence, nicoletti2024information} and biological sciences \cite{nejatbakhsh2023predicting,lee2022predicting,mu2024mendelian, lin2018temporal}, has significantly expanded the scope of their applications. For example, the structure of citation networks in scientific research is governed by the causal dependencies between ideas, where innovations are inspired by prior knowledge \cite{redner1998popular}. Similarly, in biological evolution, the inheritance of traits follows causal dependencies from existing traits \cite{lin2018temporal,mu2024mendelian}.

A causal network, from a physics perspective, is inherently rigid due to the immutability of past events. Vertices represent events, and directed edges denote fixed causal relationships. Unlike general networks, causal networks cannot be rewired or retroactively modified without violating temporal and causal consistency. This rigidity can be represented by directed acyclic graphs (DAGs), 
reflecting the unidirectional arrow of time. Consequently, causal networks serve as static records of chronological and causal relationships, distinguishing them from dynamic, modifiable general networks \cite{krapivsky2001degree, albert2002statistical, zhou2023nature}. For example, while general networks like social media follower-followee relationships can evolve dynamically via rewiring or deletion, causal networks provide static records of interactions such as ``follow" or ``unfollow," preserving the historical flow of actions. In causal networks, such dynamics are simplified through an event creation mechanism, leveraging the growth network approach \cite{barabasi1999emergence, bianconi2001bose, golosovsky2012stochastic, kharel2022degree, jin2019emergence}, enabling broader applicability.

Causal networks are characterized by various correlations that quantify the dependencies between entities. Of particular importance are topological correlations, which reveal the non-random organization of the network structure \cite{zheng2024topological,maslov2002specificity,park2003origin,van2010influence}. These correlations provide critical insights into how causal relationships shape the network architecture. Empirical evidence increasingly suggests that topological correlations are crucial for the development of causal networks \cite{hidalgo2021economic, park2023papers,jin2019emergence}. For instance, topological correlations in citation networks are believed to govern the inheritance patterns of scientific impacts \cite{kuhn2014inheritance}. Similarly, in ecological event networks, topological correlations that reflect the causality between different events play a key role in shaping microbial community structures \cite{delogu2024forecasting}. While there has been substantial progress in quantifying topological correlations for general networks \cite{newman2002assortative, newman2003social, park2003origin, pastor2001dynamical}, relatively few studies have attempted to model network correlations explicitly \cite{caldarelli2002scale, servedio2004vertex, goh2001universal}. Moreover, these models primarily focus on non-causal networks. The fundamental distinction in the causal network implies that traditional network correlation models, particularly those relying on dynamic rewiring mechanisms \cite{krapivsky2001degree, albert2002statistical, zhou2023nature}, cannot fully capture the unique correlation patterns that arise within causal networks.

In this letter, we propose a general correlated growth framework for casual networks, incorporating two key mechanisms: 1) {\it Dynamic correlation}, which captures the long-range temporal correlations during the individual degree growth; 2) {\it Causal correlation}, where causal relationships between individuals and their successors are transmitted across generations. By analytically solving the joint degree distribution, we explicitly demonstrate that degree correlations emerge from marginal dependencies on both dynamic and causal correlations. Applying our theory to large-scale innovation networks across four disciplines, we find strong quantitative agreement between our theoretical predictions and empirical observations. As a unified approach to modeling correlated growth in causal networks, our framework has the potential to impact a wide range of fields.

{\it Theoretical Framework.} 
Consider a causal network model that starts with $n_0$ events at time $t=0$. As the network evolves, each newly added event $i$ connects to $m_i$ preexisting events introduced at earlier times, yielding a DAG that ensures causality. Additionally, each event $i$ is assigned a state $\s_i$, which may include both discrete and continuous attributes. For instance, in a social network, $\s_i$ could encode the type of interaction (discrete) alongside its intensity or duration (continuous). To capture a wide range of scenarios, we divide $\s_i$ into two subspaces: an observed component $\s_{\mathrm{obs}}$, reflecting publicly accessible properties, and a latent component $\s_{\mathrm{lat}}$, encoding hidden variables that influence network evolution. 

Once an event with state $\s$ is created at time $t$, its causal relationships can continue to grow. We capture this individual-level growth via a Green’s function $G_{\s}(k,\tau | k',\tau')$, which gives the probability that an event with $k'$ connections at age $\tau'$ grows to $k$ connections at a later age $\tau$. Here, $\tau' \ldef t' - t_i$ and $\tau \ldef t - t_i$, measured from the event’s creation time $t_i$. The corresponding mean accumulative connection, $\bark(\s, \tau) \ldef \sum_k k G_\s(k,\tau|0,0)$, then tracks how an event’s expected connections evolve over its lifetime. As an example, a Poisson process with rate $\lambda$ yields $G_\lambda(k,\tau|k',\tau') = \frac{\left[\lambda (\tau - \tau')\right]^{k - k'} e^{-\lambda (\tau - \tau')}}{(k - k')!}$, and $\bark(\s,\tau) = \lambda\tau$. However, for a process with temporal correlation, the corresponding Green’s function  $G$  may exhibit a fat tail rather than an exponential decay. In our framework, $G$ remains a general input to accommodate different growth processes. 

Global network growth arises from the collective growth of individual events. To describe this process, we focus on $n(\s,t)$, the rate at which new events with state $\s$ appear. The directed edges from newly introduced vertices to existing ones encode causal correlations; for example, an ``unfollow'' action can only occur after a corresponding ``follow.'' To capture these correlations, we employ a mean-field causal kernel $K(\s,t| \s',t')$, which gives the probability that a vertex created at time $t$ has state $\s$ given that it links to a vertex with state $\s'$ created at an earlier time $t'<t$. The event-creation rate then satisfies
\begin{align}\label{eq:master}
&n(\s,t) = n_0(\s)\delta(t) + \notag\\
&\frac{1}{\avek}\int_0^{t}\d t'\int \d \s' \; n(\s',t')K(\s,t|\s',t')\partial_t\overline k(\s',t-t'),
\end{align}
where $\langle k \rangle$ is the average degree. For simplicity, in the discussion below, we consider the case of a fixed outdegree, $m_i=m$, so that $\langle k \rangle = m$. In more general cases, $m_i$ can be incorporated into the state $\s_i$. 

To solve Eq. \eqref{eq:master}, we focus primarily on its stationary solution. In this regime, $n(\s,t)$ factorizes asymptotically as $n(t)\psi(\s)$, where $\psi(\s)$ is the stationary state distribution, and 
\(
n(t) \equiv \int n(\s,t)\,\d \s \;=\; \frac{\d N(t)}{\d t}
\)
is the total growth rate of the network size. One can show that the existence of such a stationary solution requires the growth process to exhibit time-translation invariance (TTI), i.e.\ $K(\s,t| \s',t') = K(\s|\s', t-t')$ and $n(t)/n(t') \sim f(t-t')$, so that only time differences matter\citeSM{1.2}. Under these conditions, we predict an exponential growth
\begin{equation}\label{eq:nt}
    n(t)\,\sim\,e^{rt},
\end{equation}
where $r$ is the network growth rate. The analysis above also implies the asymptotic state-age distribution,  
\begin{equation}\label{eq:Psi}
    \Psi(\s, \tau) \ldef \lim_{t\to\infty}\frac{n(\s,t-\tau)}{\int_0^t n(\s, t-\tau) d\tau} = \psi(\s) r e^{-r\tau},
\end{equation}
factorizes into the product of the state and age distributions.

Substituting Eq.~\eqref{eq:nt} into Eq.~\eqref{eq:master} and taking $t\to\infty$ yields leads the stationary master equation
\begin{equation}\label{eq:stationary}
  \psi(\s)=\frac{1}{\avek}\int_0^\infty \d \tau\int\d \s' K(\s|\s',\tau)\partial_\tau {\bark(\s',\tau)} e^{-r\tau}\psi(\s'), 
\end{equation}
subject to $\int \d \s\,\psi(\s)=1$. Given the causal kernel $K$ and the individual growth Green's function $G$, which defines $\bark$, this self-consistent equation determines both $r$ and $\psi(\s)$ in the thermodynamic limit, along with all physical observables to be discussed. 

{\it Network Characteristics.} Once $\psi(\s)$ and $r$ are obtained from Eq.~\eqref{eq:stationary}, the state-age distribution $\Psi(\s,\tau)$ can be determined using Eq.~\eqref{eq:Psi}, enabling an analytical computation of various network characteristics. We now outline the computation of the degree distribution $P(k)$, which represents the probability of finding a vertex with in-degree $k$, and the joint degree distribution $P(k,k')$, which gives the probability of finding an edge from a vertex with in-degree $k$ to a vertex with in-degree $k'$. The latter is a fundamental measure that characterizes network correlations.

A key quantity in our framework for understanding these network characteristics is the joint state-age distribution,
\begin{align}
    \mathbf{\Psi}(\s,\tau; \s',\tau') \ldef \frac{1}{\avek} K(\s|\s',\tau'-\tau)\partial_{\tau'} {\overline{k}(\s',\tau'-\tau)} \Psi(\s',\tau'),\notag
\end{align}
which quantifies the probability of finding an edge from a vertex with state $\s$ and age $\tau$ to a vertex with state $\s'$ and age $\tau'$ ($\tau' > \tau$). Using Eq.~\eqref{eq:stationary}, one obtains the marginal probabilities at either of these two vertices:
\(
\int \mathbf{\Psi}(\s,\tau; \s',\tau')\d\s'\d\tau' = \Psi(\s,\tau)
\)
and
\(
\int \mathbf{\Psi}(\s,\tau; \s',\tau')\d\s\d\tau = \bark(\s',\tau')\Psi(\s',\tau')/\avek.
\)

Since $G_{\s}(k,\tau|0,0)$ gives the probability that a vertex with state $\s$ and age $\tau$ attains in-degree $k$, the degree distribution is given by
\(
P(k)=\int \d\s\int_0^{\infty}\d \tau\; G_{\s}(k,\tau|0,0)\Psi(\s,\tau),
\)
which can be interpreted as the averaged Green's function,
\begin{equation}\label{eq:Pk}
    P(k) = \ave{G_\s(k,\tau|0,0)},
\end{equation}
where the average $\ave{\ldots} \ldef \int \ldots \mathbf{\Psi}(\s,\tau; \s',\tau') \d\s \d\tau \d\s'\d\tau'$ is taken over the joint state-age distribution. Similarly, one finds\citeSM{1.3}
\begin{align}
    P(k,k') = \ave{\frac{k'}{\overline{k}(\s',\tau')} G_{\s'}(k',\tau'|0,0) G_\s(k,\tau|0,0)},
\label{eq:pkk}
\end{align}
where $k'$ represents the number of edges connecting to the vertex, and $\overline{k}(\s',\tau')$ ensures that $\frac{k'}{\overline{k}(\s',\tau')} G_{\s'}(k',\tau'|0,0)$ is properly normalized. Equation~\eqref{eq:pkk} implies that $k$ and $k'$ are conditionally independent given the state and age, with their overall correlation mediated by the marginal dependencies in the joint state-age distribution. In other words, degree correlation is entirely encoded in the state-age correlation within $\mathbf{\Psi}$.


{\it Empirical Validation.}
To validate our theory against empirical data, we applied it to the scientific innovation system, which can be understood as a causal network, where the creation of innovations causally depends on existing knowledge. We used the Web of Science (WOS) dataset spanning 1970--2014, comprising 44M publications and 800M citations. For simplicity, the causal network considered here is the citation network, focusing on events of paper publication with a single type of causal relationship $\s_\text{obs} =$ ``cite", which links new knowledge to its intellectual predecessors. Using the WOS dataset, we extracted four citation networks across different disciplines, including biology, chemistry, mathematics, and physics, comprising 13M, 9M, 3M, and 7M vertices, respectively \citeSM{2.1}.

\begin{figure}
  \includegraphics[width=1\linewidth]{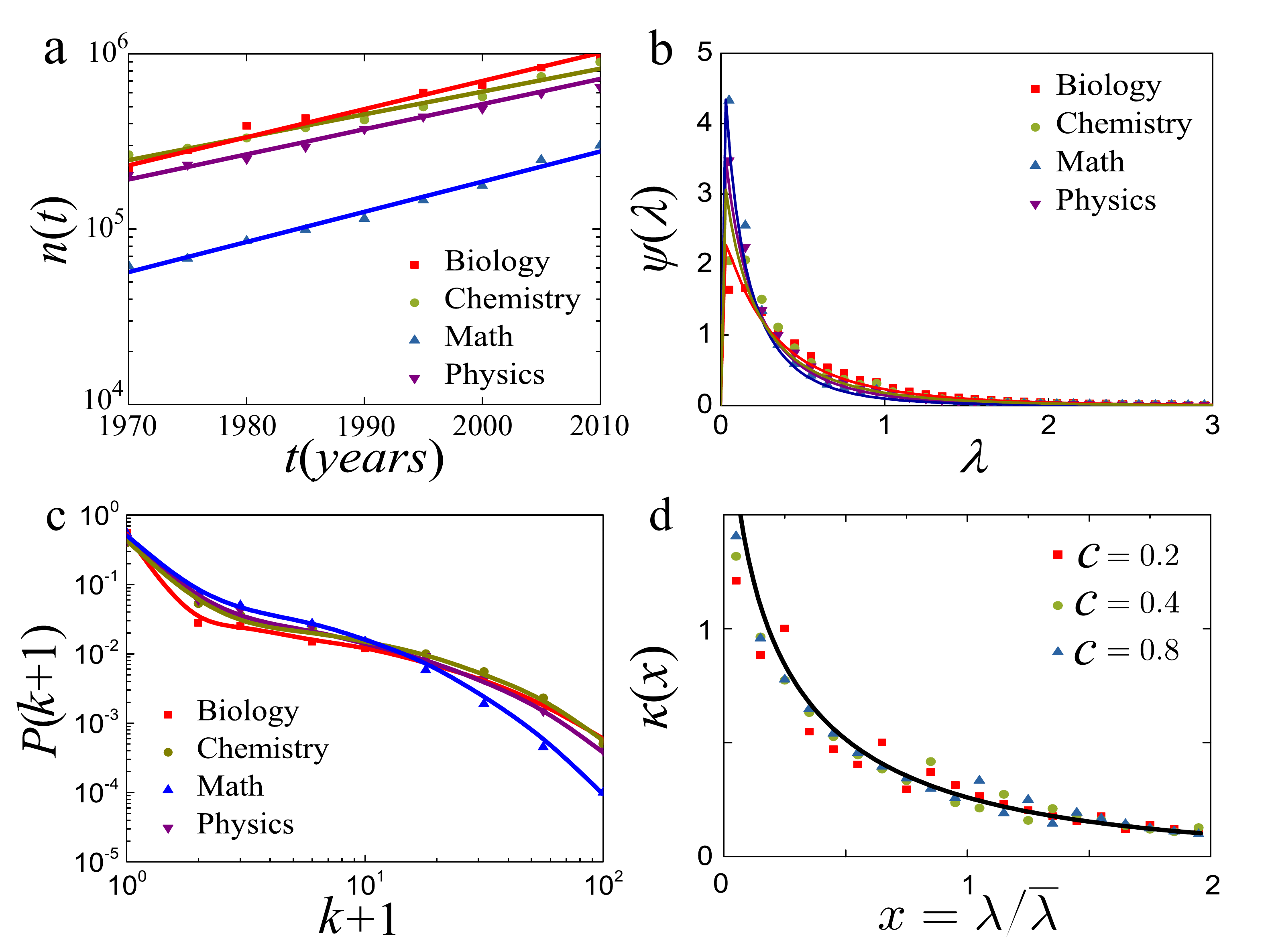}
  \caption{Empirical results (scatter points) vs. theoretical predictions (solid lines) for  
(a) The network size $n(t)$ of citation graphs across biology, chemistry, mathematics, and physics. Theoretical predictions are shown as straight lines with predicted growth rates.  
(b) The probability distribution of states $\psi(\s)$.  
(c) The degree distribution $P(k)$.   
(d) The empirically measured causal correlation kernel $K(\lambda|\lambda', \tau)$ for physics citation graphs, rescaled by its average value $\overline{\lambda}(\lambda',\tau)$. The rescaled kernel collapses onto a master curve. Similar plots for other subjects are provided in the Supplementary Material.}
  \label{fig:growth}
\end{figure}

\begin{figure}
    \includegraphics[width=1\linewidth]{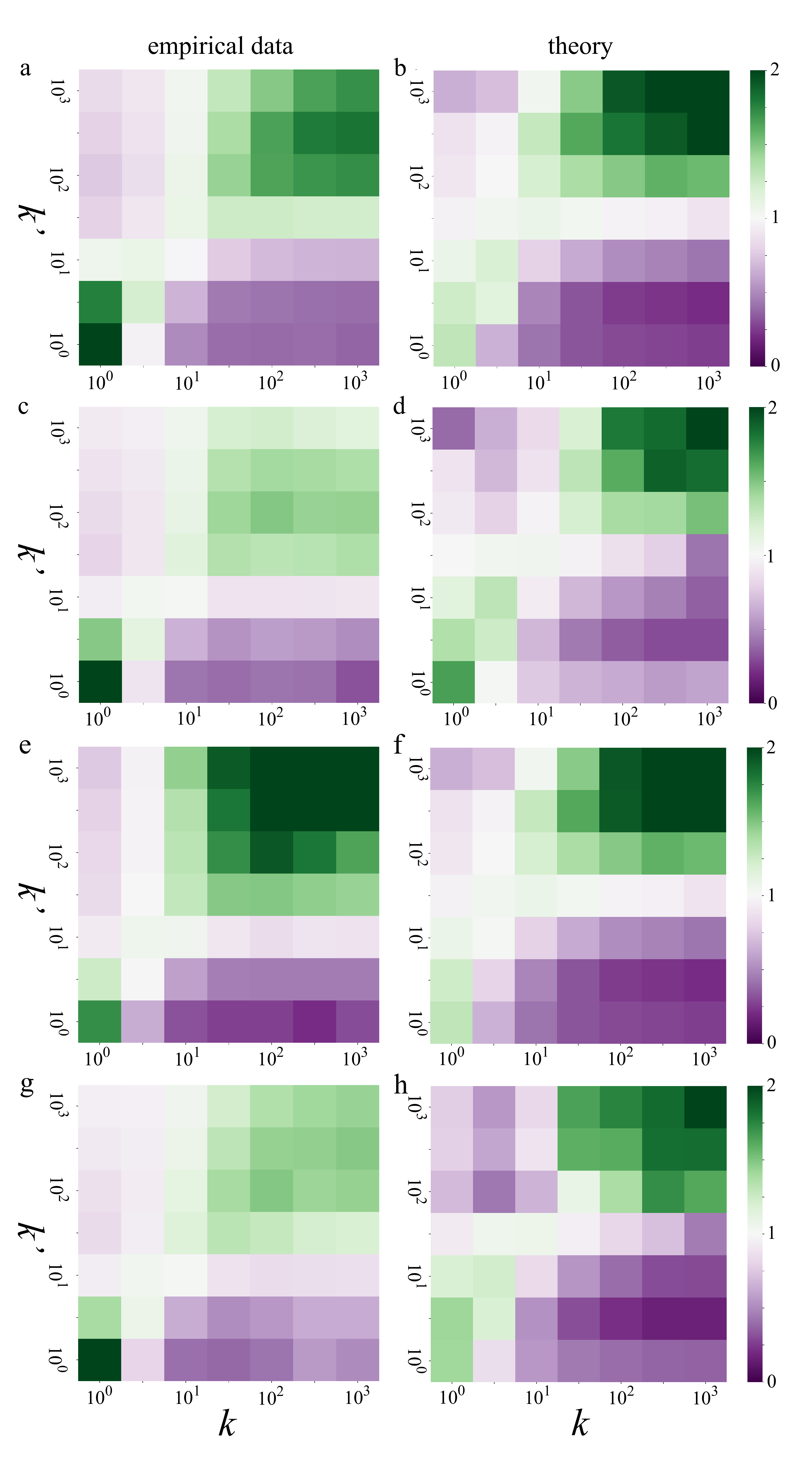}
    \caption{Normalized joint degree distribution for (a–b) biology, (c–d) chemistry, (e–f) mathematics, and (g–h) physics. The left and right columns represent empirical measurements and theoretical predictions, respectively.}
    \label{fig:pkk}
\end{figure}

Existing studies suggest that the growth process at the individual level follows a reinforced Poisson process (RPP), incorporating both preferential attachment and aging \cite{shen2014modeling,wang2013quantifying}. Specifically, the growth rate is given by $ \Lambda{(k,\tau)}=\lambda \phi(\tau)(k+k_0)$, where $\lambda$ represents fitness. The aging function, typically modeled as a log-normal distribution $\phi(t)=\frac{1}{\sqrt{2\pi}\sigma t}\exp\left[-\frac{(\ln{t}-\mu)^2}{2\sigma^2}\right]$, captures the long-tail dynamic correlation. The corresponding Green's function is given by the negative binomial distribution\citeSM{1.1}, with the average growth 
\begin{equation}\label{eq:avek}
\bark(\lambda,t)=k_0\left(\exp\left(\lambda\int_{0}^{t}\phi(t')\d t'\right)-1\right),
\end{equation}
which implies a finite ultimate degree $\bark_\infty(\lambda) \ldef \bark(\lambda,\infty) = k_0 \left(e^\lambda-1\right)$. While Eq. \eqref{eq:avek} successfully captures individual citation growth, it has several drawbacks: (1) it does not account for causal correlations, and (2) it assigns the parameters $\lambda$, $k_0$, $\mu$, and $\sigma$ to each paper, leading to a large number of free parameters that are experimentally determined. In particular, the number of parameters scales with the number of vertices, $O(N)$. Such limitations are well-documented in fitness models \cite{barabasi1999emergence,bianconi2001bose}.

Our theory addresses the above drawbacks by treating individual parameters as latent states, $\s_\text{lat}$. This approach allows our theory \eqref{eq:stationary} to predict the distribution of these latent states, thereby avoiding the large parameter space. To solve Eq.~\eqref{eq:stationary}, it is necessary to determine the causal correlation kernel, $K$. For simplicity, we focus solely on the fitness parameter $\s_\text{lat} = \lambda$, keeping all other parameters $\sigma = 1$ and $\mu = 2.3$ fixed globally. Figure~\ref{fig:growth}d plots a rescaled $K(\lambda|\lambda',t)$, finding that it collapses into master curves after scaling by its mean $\overline{\lambda}$, as  
\begin{align}\label{eq:kappa}
    K(\lambda|\lambda', t)=\frac{1}{\overline{\lambda}(\lambda',t)}\mathcal{K}\left(\frac{\lambda}{\overline{\lambda}(\lambda',t)}\right).
\end{align}  
Here, the mean fitness $\overline{\lambda}(\lambda',t)$ is an exponential function of the ratio $\bark(\lambda', t)/ \bark_\infty(\lambda')$, as shown in our previous work \cite{liu2023correlated}. Moreover, the universal distribution $\mathcal{K}$ is best fitted by the mixed Weibull distribution function \citeSM{2.3}.

Substituting Eqs.~(\ref{eq:avek}--\ref{eq:kappa}) into Eq.~\eqref{eq:stationary}, our theory predicts the network growth rates $r \approx 0.037$, $0.03$, $0.04$, and $0.028$ for biology, chemistry, mathematics, and physics, respectively, which are in excellent agreement with the empirical measurements (Fig.~\ref{fig:growth}a). Figure~\ref{fig:growth}b compares the predicted fitness distribution $\psi(\lambda)$ with the empirical data, showing remarkable consistency. Here, the empirical $\psi(\lambda)$ is obtained by fitting the RPP model to individual papers. Additionally, the degree distribution $P(k)$ is computed analytically using Eq.~\eqref{eq:Pk}, again aligning perfectly with experimental results (Fig.~\ref{fig:growth}c).

To validate the model's prediction of topical correlation, we calculate the joint degree distribution $P(k,k')$, using Eq.~\eqref{eq:pkk}. Figure \ref{fig:pkk} plots the normalized joint degree distribution $R(k,k') \ldef \frac{P(k,k')}{P_{rand}(k,k')}$, i.e., $P(k,k')$ compared to the randomized one \cite{maslov2002specificity}. Our results show that the theoretical prediction successfully captures the assortative nature of the citation graph, aligning well with empirical measurements. 

\begin{figure}
  \includegraphics[width=1\linewidth]{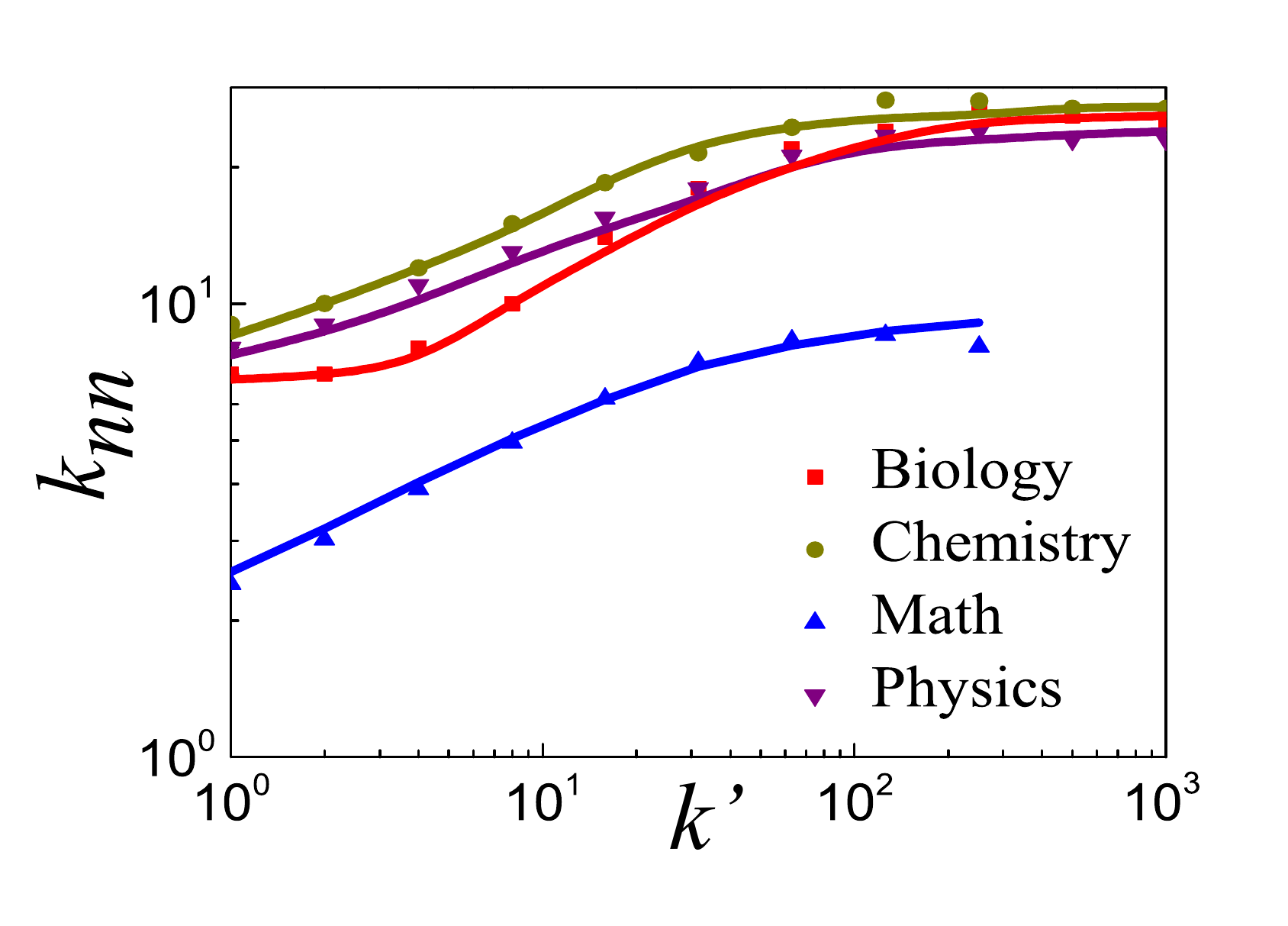}
  \caption{K-nearest-neighbor $k_{nn}(k')$ for both empirical measurements (scatter points) and theoretical predictions (solid curves) across four subjects. 
}
  \label{fig:knn}
\end{figure}

To compare our theory with empirical data more quantitatively, we further calculate the $k$-nearest-neighbor function, defined as $k_{nn}(k') \ldef \sum_{k}kP(k|k')$, which measures the dependency of a vertex's degree on that of its neighbors. Based on Eq.~\eqref{eq:pkk}, we obtain\citeSM{1.4}
\begin{align}\label{eq:knn}
   k_{nn}(k')= \left\llangle\overline k(\lambda,\tau)  \right\rrangle,
\end{align}
where $\llangle\ldots\rrangle $ denotes the average over the conditional distribution $P(\lambda, \tau; \lambda', \tau'|k')$, defined by $\overline{k}(\lambda',\tau')^{-1} G_{\lambda'}(k',\tau'|0,0) \mathbf{\Psi}(\lambda,\tau;\lambda',\tau')$, up to a normalization factor. Interestingly, the variable $k'$ does not explicitly appear in the formula but is instead implicitly encoded by the distribution $P(\lambda, \tau; \lambda', \tau'|k')$. Thus, as $k'$ varies, the underlying distribution changes, effectively leading to different expected values. The scatter plot in Fig.~\ref{fig:knn} shows that the empirical $k_{nn}$ increases with $k'$ and converges to finite values for large $k'$, indicating a clear positive correlation. Our theoretical prediction \eqref{eq:knn}, represented by the solid curves, exhibits good agreement across all four networks. Notably, our theory suggests that the saturation observed for large $k'$ is not a finite-size effect but rather a consequence of both the causal kernel in Eq.~\eqref{eq:kappa} and the finiteness of the ultimate degree $\bark_\infty$\citeSM{1.5}.

Moreover, the joint degree distribution $P(k,k')$ enables the analytical computation of the assortativity coefficient 
\(
r_{corr}\ldef\frac{Cov(k,k')}{\sqrt{Var(k) Var(k')}}
\)  \cite{newman2002assortative}. Using the law of total covariance and the marginal independence of $k$ and $k'$, Eq.~\eqref{eq:avek} naturally leads to 
\citeSM{1.6}
\begin{align}
   r_{corr} =  \frac{Cov\left(\bark(\lambda,\tau),\bark(\lambda',\tau')\right)}{\sqrt{Var\left(\bark(\lambda,\tau)\right) Var\left(\bark(\lambda',\tau')\right)}},
\end{align}
as the correlation coefficient over $\mathbf{\Psi}$. Table~(\ref{tab:rcorr}) summarizes the assortativity $r_{corr}$ for both empirical results and theoretical predictions, demonstrating good agreement.

\begin{table}[htbp]
\centering
\caption{Assortativity coefficients}
\label{table1}
\begin{tabular}{ccc}
Subjects & Experiment & Theory\\
\hline
Biology & $0.20\pm 0.02$ & $0.18$\\
Chemistry & $0.11\pm 0.01$ & $0.14$\\
Math & $0.18\pm 0.01$ & $0.18$\\
Physics & $0.11\pm 0.02$ & $0.13$\\
\end{tabular}\label{tab:rcorr}
\end{table}

{\it Discussion.} Our study demonstrates that topological correlations in causal networks naturally originate from fundamental causal and dynamical correlations during network growth. Analytical results explicitly reveal that these correlations induce marginal dependencies, which in turn generate observable degree correlations. We validate our theoretical framework by applying it to empirical citation graphs, which serve as representative real-world examples.

Compared to existing fitness-based models, which assign a separate fitness value to each vertex and require $O(N)$ parameters for a network with $N$ vertices, these models often suffer from overfitting, thereby limiting predictive power. In contrast, our model introduces causal dependencies between fitness values, encoded in a causal kernel, which reduces the number of free parameters from $O(N)$ to $O(1)$. This structured approach mitigates overfitting while improving both explanatory and predictive capabilities. In this way, our method addresses notable limitations of existing models \cite{wang2013quantifying,liu2018hot}, demonstrating that causal dependencies provide a more compact and generalizable framework for modeling real-world networks.

While our theory successfully captures key causal dependencies, several open questions remain. One major challenge is the incorporation of time-varying state distributions, particularly in causal networks exhibiting non-stationary growth. If Equation \eqref{eq:stationary} admits no solution, stationarity breaks down. This can occur when the causal kernel has sufficiently broad tails. Consequently, the network undergoes super-exponential growth, violating TTI. This scenario is closely related to the ``winner-takes-all'' phase, where a small fraction of vertices accumulate a disproportionate share of influence, analogous to Bose-Einstein condensation \cite{bianconi2001bose}. Since this behavior does not arise in the empirical data we analyze, we defer its detailed investigation to future work. Another important direction is to explore topological correlations beyond the nearest-neighbor level. Higher-order structural dependencies may reveal additional causal mechanisms governing network formation, further enriching our understanding of dynamical correlations in complex systems.

Despite these open challenges, our findings provide a robust foundation for understanding correlations in causal networks. By incorporating causal correlation kernels, our framework applies to a wide range of systems, including social media, biological evolution, and economic growth. This approach offers new insights into the interplay between causal structure and network topology, with broad implications across physics, economics, social sciences, and biology.

\end{document}